\documentclass[prl,aps,epsfig,twocolumn,showpacs]{revtex4}
\usepackage{epsfig,amssymb}

\newcommand\beq{\begin{equation}}
\newcommand\eeq{\end{equation}}
\newcommand\bea{\begin{eqnarray}}
\newcommand\eea{\end{eqnarray}}
\newcommand\non{\nonumber}

\newcommand\al{\alpha}
\newcommand\be{\beta}

\newcommand\la{\lambda}
\newcommand\om{\omega}

\newcommand\tv{\tilde v}
\newcommand\dg{\dagger}

\newcommand\bib{\bibitem}

\begin{document}

\textheight=23.8cm

\title{\Large Inter-edge interactions and novel fixed points at a junction 
of quantum Hall line junctions}
\author{\bf Sourin Das$^1$, Sumathi Rao$^2$ and Diptiman Sen$^3$} 
\affiliation{
\it $^1$ Department of Condensed Matter Physics, The Weizmann 
Institute of Science, Rehovot 76100, Israel 
\\
\it $^2$
Harish-Chandra Research Institute, Chhatnag Road, Jhusi, Allahabad 
211019, India \\
$^3$ Centre for High Energy Physics, Indian Institute of Science, Bangalore 
560012, India.}
\date{\today}

\begin{abstract}

We show that novel fixed points (characterized by matrices which specify the
splitting of the currents at the junction) can be accessed in a system which
contains a junction of three quantum Hall line junctions. For such a junction 
of fractional quantum Hall edge states, we find that it is possible for both 
the flower (single droplet) and islands (three droplets) configurations to be 
stable in an intermediate region, for a range of values of the inter-edge 
repulsive interactions.
A measurement of the tunneling conductance as a function of the gate voltage 
controlling inter-edge repulsions can give a clear experimental signal of 
this region.

\end{abstract}
\maketitle
\vskip .6 true cm

Line junctions \cite{renn} between the edge states of a fractional quantum 
Hall system \cite{wen} allow the realization of one-dimensional systems of 
interacting electrons with a tunable Luttinger parameter \cite{gogolin}. A 
line junction is formed by creating a narrow barrier which divides a fractional
quantum Hall liquid (FQHL) such that there are chiral edge states flowing in 
opposite directions on the two sides of the barrier \cite{kang,roddaro};
the edges interact with each other through Coulomb repulsion. A line junction
is thus similar to a non-chiral quantum wire; however, the physical separation
between the two edges of the effective non-chiral wire allows for a greater 
control over the strength of the interaction between them.

Recent experiments have shown that the geometry of the quantum Hall droplet 
and the location of the points across which tunneling occurs can influence the
degree of back-scattering and therefore the transport. Motivated by this, we 
will study here a FQHL droplet with three narrow barriers as shown in Fig. 1.
(Junctions of three quantum Hall edges have been studied earlier 
\cite{chen,oshikawa,chamon}, but not in the context of line junctions). The 
width of the narrow barrier between the edges can be tuned to control the 
Coulomb repulsion between the two edges on its opposite sides; this in turn 
controls the Luttinger parameter $g$ in each non-chiral wire which is formed 
by the two edges. Unlike the typical split Hall bar model, this geometry 
offers access to a new class of tunnelings and fixed points. When there is 
perfect symmetry between the three barrier gates, we find that there is a range
of $g$ for which both the flower fixed point (fully disconnected in terms of 
wires) and the islands fixed point (chiral in terms of wires) are stable. We 
compute the scaling of the tunneling conductances around these fixed points.

\begin{figure}[htb]
\begin{center}
\epsfig{figure=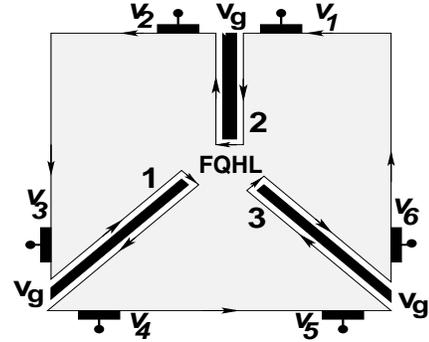,width=5.8cm,height=4.7cm}
\vspace{-0.3cm}
\end{center}
\caption{Single droplet (flower configuration) of FQHL. Line junctions are
formed by the gate voltages $V_g$. $V_i$ denote the potentials which drive 
currents between different edges.}
\end{figure}

These fixed points are obtained by imposing boundary conditions on the currents
via a matrix which splits the currents into the three `wires'. Although many 
consistent (conformally invariant) boundary conditions are possible 
\cite{nayak,chen,oshikawa,chamon}, we will focus on certain simple boundary 
conditions which can be visualized in terms of processes involving the 
electrons and quasi-particles (quasi-electrons and quasi-holes) at the 
junction. 

The Lagrangian for a system of three quantum Hall line junctions is given by
\bea
L &=& {1\over 4\pi} ~\sum_{i=1}^3 ~[~ \int_0^\infty dx ~\partial_x \phi_{iO}~
(-\partial_t -v\partial_x) ~\phi_{iO} \non \\
&& ~~~~~~~~~~~+ \int_{-\infty}^0 dx ~\partial_x \phi_{iI} ~(-\partial_t - v 
\partial_x) ~\phi_{iI} ~] \non \\ 
&& +~ {v \la \over \pi} ~\sum_{i=1}^3 ~\int _0^\infty dx ~\partial_x 
\phi_{iO}(x) ~\partial_x \phi_{iI}(-x) ~,
\label{lag1}
\eea
where $v$ denotes the velocity, $i$ labels the wire, the incoming fields
$\phi_{iI}$ are defined from $x=-\infty$ to $0$, and the outgoing fields 
$\phi_{iO}$ from $x=0$ to $\infty$. The geometry allows for a screened Coulomb
interaction between the left and right movers with a strength $\la$ which has 
to be positive; $\la$ can be varied by a gate potential.
When the gate potential is large, the left and right movers are well-separated
and $\la$ is small; when it is small, the two modes move closer to each
other and $\la$ is large. We restrict ourselves here to the case where there 
is no hopping between the modes. Note that $\la$ is related to the parameter 
$g$ of a non-chiral Luttinger wire as $g =[(1-\la)/(1+\la)]^{1/2}$.
We therefore choose $\la$ to be less than one.

The quasi-electron and electron operators are given by $\psi_{qe} = \eta_i 
e^{i{\sqrt \nu} \phi_i}$ and $\psi_{el} = \chi_i e^{i\phi_i /{\sqrt \nu}}$ 
respectively, where $\nu ~(=1/3, 1/5, \cdots)$ is the FQHL filling, and 
$\eta_i$ and $\chi_i$ are the Klein factors for quasi-electrons and 
electrons respectively. The density fields canonically conjugate to $\phi$ 
are given by $\rho_{i,I/O} = -(1/2\pi) \partial_x \phi_{i,I/O}$, so that 
\bea
{}[\phi_{iI/O} (x), \rho_{jI/O} (y)] &=& \delta_{ij} ~\delta (x-y) ~~ 
{\rm for} ~~ x, y </> 0 ~, \non \\
{}[\phi_{iI} (x), ~\rho_{j0} (y)] &=& 0 ~~ {\rm for} ~~ x < 0,~ 
y > 0 ~.
\label{comm}
\eea

At the junction, the Lagrangian in Eq. (\ref{lag1}) must be supplemented by 
boundary conditions which ensure that the current (given by $j_{i,I/O} =
(1/2\pi) \partial_t \phi_{i,I/O}$) is conserved, and that Eqs. 
(\ref{comm}) are satisfied. This implies that the fields must be 
related at the junction as ${\vec\phi}_{O} = S {\vec\phi}_{I}$,
where the $3\times 3$ splitting matrix $S$ is real and orthogonal, and each of
its columns (or rows) add up to 1. The latter conditions ensure that the fields
satisfy $\sum \phi_{iO} = \sum_i \phi_{iI}$, so that the current is conserved 
at the junction. 

We now consider some simple forms of $S$, which are the identity matrix $I$ 
and the two chiral matrices, namely, $M_+$ with $M_{13}=M_{21}=M_{32}=1$ and 
all the other matrix elements equal to zero,
and $M_- = M_+^T$. For a given sign of the magnetic field, only one chirality 
is possible, so we only consider one of them, say, $M_+$. 
We will consider a given value of the FQHL filling $\nu < 1$, and study the 
scaling dimensions of various tunneling operators as functions of $\la$ or 
the Luttinger parameter $g$.

The case $S=I$ corresponds to the situation in Fig. 1, in which current from 
the incoming edge $i$ goes entirely to the outgoing edge $i$. Since there is 
only one droplet, one can consider both electron and quasi-particle tunneling 
between two edges, say, between the incoming edge 1 and the outgoing edge 2.
The scaling dimensions of this operator can be computed after performing
a Bogoliubov diagonalization given by $\phi'_{O/I} =
[(1+g)\phi_{iO/I} +(1-g)\phi_{I/O}]/2\sqrt{g}$ in each wire.
We find that the tunneling operator as described above has the scaling 
dimension 
$\nu /g$ for quasi-particles and $1 /(\nu g)$ for electrons. Since $\nu$ and 
$g$ are both less than 1, electron tunneling is irrelevant in the sense of the
renormalization group (RG). However, if $g > \nu$, quasi-particle tunneling is
relevant, and the configuration in Fig. 1 is unstable under an RG flow. In that
case, since tunneling between the incoming edge $i$ and the outgoing edge 
$i+1$ grows, it is reasonable to assume that the configuration in Fig. 1 flows,
at long distances, to the one in Fig. 2. [Note that in the absence of Coulomb
interaction between the edges, $g=1$ is greater than $\nu$; hence the 
configuration in Fig. 1 is unstable to Fig. 2. This agrees with the usual 
expectation that a single FQHL droplet is unstable to the formation of 
multiple droplets.] 

\begin{figure}[htb]
\begin{center}
\epsfig{figure=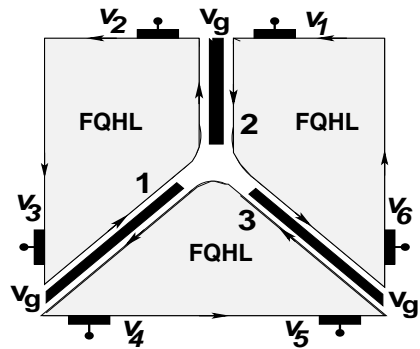,width=5.8cm,height=4.7cm}
\vspace{-0.3cm}
\end{center}
\caption{Three droplets (islands configuration) of FQHL. The gate voltages and
potentials are defined as in Fig. 1.}
\end{figure}

The case $S=M_+$ corresponds to Fig. 2. In this case, only electrons can 
tunnel between, say, the incoming edge 1 and the outgoing edges 1 or 3; the 
conservation of charge (in integer multiples of an electron) in the 
individual droplets prevents tunneling of quasi-particles from the incoming 
edge 1 to the outgoing edges 1 and 3. To calculate the scaling dimension of 
the tunneling operator, we first carry out the Bogoliubov diagonalization in 
each wire and then rewrite the boundary condition in terms of the free 
incoming and outgoing fields, i.e.,
\beq
{\vec\phi'}_O ~=~ \frac{(1+g)S +(1-g)I}{(1+g)I+(1-g)S} ~{\vec\phi'}_I ~.
\eeq
The scaling dimension of the electron tunneling operator between any incoming 
edge and outgoing edge is then found to be $4g /[\nu (3 + g^2)]$. Note that we
reproduce the scaling dimensions obtained in Refs. \cite{oshikawa,chamon} near
the chiral fixed points, without using Klein factors or mapping to the 
dissipative Hofstader model.
This is because we compute the scaling dimension of weak tunneling directly
at the islands fixed point of Fig. 2, rather than studying the strong 
tunneling limit (with multiple hoppings involving Klein factors) of the flower
fixed point of Fig. 1. Thus we identify the islands configuration (and its
time-reversed form) with $\chi_{\pm}$ \cite{oshikawa,chamon}. (For $g$ close to
1, these reduce to the chiral fixed points first studied in Ref. \cite{lal}).

We find that the dimension of the electron tunneling operator at the chiral 
fixed point is less than 1 if $g< g_c$, where
$g_c = \frac{2}{\nu} - {\sqrt {\frac{4}{\nu^2} - 3}}$;
this is equal to $0.255$ for $\nu=1/3$ (this value of $g$ corresponds to
$\la = 0.877$). Hence the configuration in Fig. 2 is unstable if $g< g_c$ and 
stable if $g > g_c$. For $g < g_c$, since tunneling between the incoming 
edge 1 and the outgoing edge 1 grows, it is reasonable to assume that Fig. 2 
flows under RG to Fig. 1. We thus see that the flower in Fig. 1 is stable if 
$g < \nu$, and the islands in Fig. 2 is stable if $g > g_c$. Since $g_c$ is 
less than $\nu$ (for $\nu < 1$), we have the interesting situation that in 
the intermediate range $g_c < g < \nu$, the configurations in Figs. 1 and 2 
are both stable; this implies that there must be an unstable fixed point lying
between the two configurations. As a function of the gate voltage controlling 
the strength of the inter-edge interactions, the single droplet is unstable to
breaking up into three droplets if the inter-edge coupling $\la < 0.877$. But 
if the gate voltage is decreased and the inter-edge interaction increases to 
$\la > 0.877$, the single droplet configuration becomes stable. These results 
are summarized in the table below.

\vspace{0.2cm}
\begin{tabular}{|c|c|c|c|c|c|}
\hline
Geo- & Tunneling & Scaling & \multicolumn{3}{c|}{RG relevance} \\
metry & Operator & dimen- & $g<g_c$ & $g_c<g$ & $g>\nu$ \\
& & sion & & ~~ $<\nu$ & \\ \hline
Flower & $e^{i(\phi_{iO}-\phi_{jI}) /{\sqrt \nu}}$ & $\frac{1}{\nu g}$ & irrel.
& irrel. & irrel. \\ \hline
Flower & $e^{i\sqrt{\nu}(\phi_{iO}-\phi_{jI})}$ & $\frac{\nu}{g}$ & irrel. & 
irrel. & rel. \\ \hline
Islands & $e^{i(\phi_{iO}-\phi_{jI}) /{\sqrt\nu}}$ & $\frac{4g} {\nu(3+g^2)}$
& rel. & irrel. & irrel.\\ \hline
\end{tabular}
\vspace{0.2cm}

One way to experimentally distinguish between the flower and islands 
configurations would be to measure the differential tunneling conductance 
$dI/dV$ between, say, the incoming edge 1 and the outgoing edge 3; the 
tunneling amplitude for this process is expected to be small in both 
configurations since those two edges are well separated. The tunneling 
conductance $G \sim b^2 V^{2(d-1)}$ where $V$ is the voltage difference (or 
temperature $T^{2(d-1)}$) for small values of $V$ (or $T$), where $d$ is the 
scaling dimension of the tunneling operator, and $b$ is the back-scattering
strength. For the flower which is stable if $g < \nu$, tunneling will be 
dominated by quasi-particles since the value of $d$ is smaller for them than 
for electrons; the exponent of $V$ (or $T$) will be given by $(2 \nu /g) -2$.
For the islands which is stable if $g > g_c$, only electrons can tunnel, and 
the exponent of $V$ will be given by $8g/[\nu (3+g^2)] - 2$. Note that the 
change from instability to stability occurs at different points for the two 
configurations, which is why there is an intermediate region where both 
configurations are stable. In Fig. 3, we plot the tunneling conductances for 
both configurations in the three regions (i), (ii) and (iii) defined in the 
caption.

\begin{figure}[htb]
\begin{center}
\epsfig{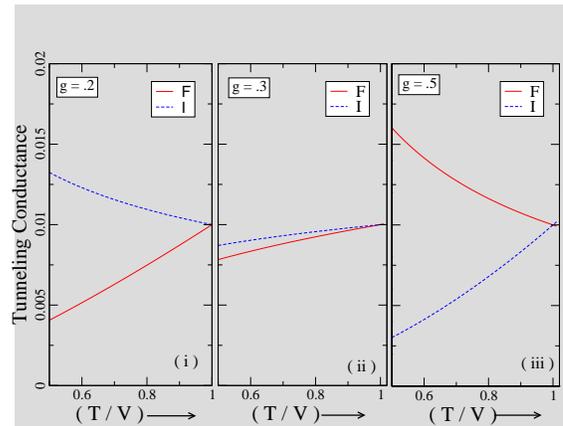}
\end{center}
\caption{Tunneling conductance in regions (i) $0<g<g_c$, (ii) $g_c<g<\nu$, and
(iii) $\nu<g<1$ as a function of the voltage or temperature, for the flower (F)
and islands (I) configurations. The quasi-particle tunneling is plotted for 
the flower, and electron tunneling for the islands. The conductance has been 
normalized to $0.01$ at the temperature $T=1$ (scaled by the cutoff temperature
$\Lambda$). The flower and islands configurations are both stable in region 
(ii).}
\end{figure}

If we start with the flower configuration with $g$ slightly less than 1 (weak
back-scattering) at high temperatures (or high voltages), and slowly reduce 
the temperature, the system flows to the islands configuration. The tunneling 
conductance at low temperatures (governed by electron tunneling) is plotted 
in Fig. 4 (line F-I, signifying that we start with the flower configuration 
at high temperatures and reach the islands configuration at low temperatures).
The experiment can be repeated after reducing $g$.
Until we reach $g=1/3$, the system always flows to the islands configuration 
at low temperatures and the tunneling conductance is governed by the F-I line. 
However, for $g<1/3$, the flower configuration is stable; 
even at low temperatures, the system remains in that configuration.
The tunneling conductance at low temperatures is governed by quasi-particle
tunneling plotted in Fig. 4 as the line F-F. Note that at $g=1/3$, the 
electron and quasi-particle tunneling operators are both marginal.

Similarly, we may start with the islands configuration at high temperatures
and look at the scaling of the conductance at low temperatures. Until we
reach $g=g_c$, the islands remains stable and the low temperature tunneling 
conductance is governed by the irrelevant electron operator (which turns 
marginal at $g_c$). If the experiment is repeated with $g<g_c$, the low 
temperature stable phase is the flower configuration, where the conductance 
is governed by the quasi-particle tunneling operator.
 
Hence, by starting with either the flower or the islands configuration at high
temperatures and changing the value of $g$ of the line junction, we should see
a dramatic change in the behaviors of the tunneling conductances at $g=\nu$ 
and $g=g_c$. {\it This is an unambiguous prediction which can be 
experimentally tested.}

\begin{figure}[htb]
\begin{center}
\epsfig{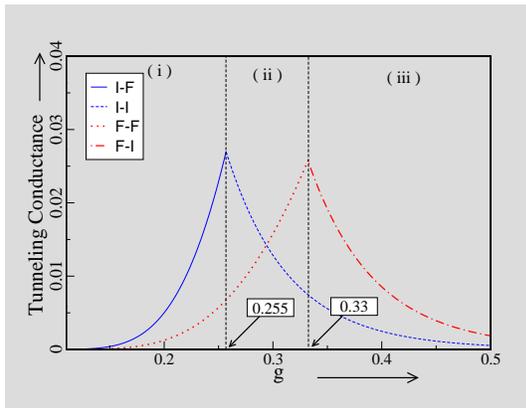}
\end{center}
\caption{Tunneling conductance as a function of $g$, starting from either the
islands configuration (I-I and I-F lines) or flower configuration (F-F and F-I
lines) at high temperature. The conductance at the marginal points has been 
normalized to be 0.025. Low temperature ($T=0.1$) conductances (quasi-particle
tunneling for the I-F and F-F lines, and electron tunneling for the I-I and 
F-I lines) have been plotted.}
\end{figure}

The three droplet and the single droplet configurations will also show 
different behaviors of the noise \cite{noise,safi}. The shot noise at the 
lowest temperatures will show signatures of both electron and quasi-particle 
hopping for the single droplet case, and a signature of only electron hopping 
for the three droplet configuration. 
The zero-frequency limit of the shot noise $S(\om)$ is proportional to the 
tunneling current $I$ and to the charge of the electron/quasi-particle which
is tunneling; the term of order $\om$ in $S(\om)$ is proportional to 
$V^{4(d-1)}$ \cite{noise}. 

For a general $S$-matrix at the boundary, we can study the problem by
solving the equations of motion following from the Lagrangian in Eq.
(\ref{lag1}); details will be reported elsewhere \cite{das2}. 
[Here, the splitting matrix and the interactions are introduced at the same 
time. 
This is different in spirit from the procedure of `delayed boundary condition'
followed in Ref. \cite{oshikawa}, where 
the boundary conditions are chosen {\it a posteriori}.]
We find that for each wave number $k$, there are three modes (labeled
by $p=1,2,3$) with the same velocity $\tv = v {\sqrt {1 - \la^2}}$. Upon 
imposing the commutation relations given in Eq. (\ref{comm}), we obtain 
\bea
&& \phi_{iI/O} (x,t) ~=~ \int_0^\infty ~\frac{dk}{\sqrt k} ~\sum_p ~
\psi_{ipI/O,k} (x,t) ~, \non \\
&& \psi_{ipI/O,k} ~=~ \al_{pk} ~(a_{ipI/O} e^{ikx} + b_{ipI/O} e^{-ikx})~ 
e^{-i\tv kt} \non \\
& & ~~~~~~~~~~~~~~~ + ~{\rm h.c.} ~, \non \\
&& {\rm with}~~ [\al_{pk} ~,~ \al_{p'k'}^\dg ] ~=~\pi ~\delta_{pp'} ~\delta 
(k - k') ~.
\eea
The wave function coefficients $a_{ip,I/O}$ and $b_{ip,I/O}$ may be compactly
written as $3 \times 3$ matrices $A_{I/O}$ and $B_{I,O}$, such that 
$(A_{I/O})_{ip} = a_{ip,I/O}$ and $(B_{I/O})_{ip} = b_{ip,I/O}$. In the 
absence of interactions, the incident waves are given by $A_I=I$, and the 
transmitted waves by $A_O = S$; the reflected waves $B_I$ and $B_O$ vanish. 
The interactions cause rescaling and reflections of the waves in each wire; 
this is governed by a parameter $\mu = {\la}/ {(1 + {\sqrt {1 - \la^2}})}$, 
which is related to the parameter $g$ as $\mu = (1-g)/(1+g)$. Furthermore, 
the boundary $S$ matrix relates the transmitted waves to the incident waves. 
We find that
\bea
A_I = \frac{I}{\sqrt {1-\mu^2}} ~, && B_I = \mu ~D ~A_I ~, \non \\
A_O = D ~A_I ~, && B_O = \mu A_I ~, 
\label{soln}
\eea
where $D = {(S - \mu I)}/{(I - \mu S)}$ is an orthogonal matrix.

We can now compute the dimension of an operator which produces tunneling at the
junction ($x=0$) between an incoming edge $i$ and an outgoing edge $j$. The 
tunneling operator is given by $O_{\be ,ij} (t) = \exp i \be (\phi_{iI} (0,t)
- \phi_{jO} (0,t))$, where $\be = {\sqrt \nu}$ and $1/{\sqrt \nu}$ for 
quasi-electrons and electrons respectively. In terms of the matrices $A$ and 
$B$ given in Eq. (\ref{soln}), the scaling dimension of $O_{ij}$ is given by
\bea
d_{\be ,ij} &=& \frac{\be^2}{2} ~\sum_p ~(A_{I,ip} ~+~ B_{I,ip} ~-~ A_{O,jp} ~
-~ B_{O,jp})^2 \non \\
&=& \frac{\be^2}{1 - \mu^2} ~[~ 1 ~-~ D_{ji} ~+~ \mu ~(~ D_{ii} ~+~ D_{jj} ~
-~ 2 \delta_{ij}~) \non \\
& & ~~~~~~~~~~ ~+~ \mu^2 ~(~ 1 ~-~ D_{ij}~) ~]~.
\label{dim}
\eea
For instance, for the electron hopping operator at the fixed point $M_+$, this
gives $d = 4g/\nu(3+g^2)$, which agrees with the earlier analysis. This 
formalism can be used to check the stability of various other fixed points 
\cite{das2}.

In summary, we have proposed a new geometry for line junctions of FQHL edges.
For $\nu=1/3$, we find that for values of the parameter $g$ (which is 
determined by the width or gate voltage of the line junction) lying in the 
range $0.255 <g< 0.333$, the single droplet (flower) and the three droplet 
(islands) phases are both stable. These phase boundaries can be experimentally
tested by measuring the voltage power law as a function of the gate voltage 
which controls $g$. 

We thank Siddhartha Lal for interesting discussions. SD acknowledges many 
stimulating and useful discussions with Yuval Gefen. SD was supported by the 
Feinberg Graduate School, Israel, and is also grateful to HRI for hospitality 
during the completion of this work. DS thanks the Department of Science and 
Technology, India for financial support under projects SR/FST/PSI-022/2000 
and SP/S2/M-11/2000.

\end{document}